\documentclass[12pt,aps]{revtex4}%
\usepackage{amssymb}
\usepackage{amsmath}
\usepackage{amsfonts}
\usepackage{graphicx}%
\providecommand{\U}[1]{\protect\rule{.1in}{.1in}}
\begin{document}
\title{Generating Anisotropic Collapse and Expansion Solutions of Einstein's Equations}
\author{E.N. Glass}
\affiliation{Physics Department, University of Michigan, Ann Arbor, MI }
\date{7 August 2013}

\begin{abstract}
Analytic gravitational collapse and expansion solutions with anisotropic
pressure are generated. Metric functions are found by requiring zero heat flow
scalar. It emerges that a single function generates the anisotropic
solutions.\ Each generating function contains an arbitrary function of time
which can be chosen to fit various astrophysical time profiles. Two examples
are provided: a bounded collapse metric and an expanding cosmological
solution.\newline\ \newline Keywords: anisotropic fluids, collapse, expansion

\end{abstract}
\maketitle

\section{INTRODUCTION}

The general problem of anisotropic fluids which undergo spherical collapse
with energy transport by heat flow was formulated some time ago by Misner and
Sharp \cite{MS65}. General relativistic models with anisotropic stress have
become increasing useful \cite{HOP08, HSW08}.\ There has been considerable
interest in anisotropic spheres \cite{BL74, CHE+81, Bay82, Bon92,Via08}
because of applications to stellar models \cite{DG02, DG04, MH03, ST12}.
Collapse has been studied in the semiclassical regime by Visser et. al.
\cite{BLSV08}. They found that Hawking radiation might prevent formation of a
trapped surface and subsequent horizon, although an exponential approach might
allow it to form in infinite time. They proposed a new class of collapsed
objects with no horizons. (In this work, the collapse example does have a
trapped surface.) Herrera and Santos \cite{HS97} have reviewed anisotropy in
self-gravitating systems. Among the topics in their extensive review are
discussions of the energy content of collapsing spheres and the stability of
perturbed solutions. Herrera et. al. \cite{HOP08} mention the possibility of a
single generating function for non-static anisotropic solutions. That
possibility is developed in this work.

In previous work \cite{Gla81}, a method of generating collapse solutions with
radial heat flow was presented. That method took a known spherical perfect
fluid solution which was either static or in shear-free collapse and mapped it
to a spherical shear-free collapse solution with radial heat flow. In this
work, a different method of constructing anisotropic solutions is presented.
The metric components are obtained by requiring zero heat flow scalar. A
single function $R(t,r)$ and a parameter "$\alpha$" generate anisotropic
solutions. Not all ($\alpha,R$) pairs are allowed, since the $\alpha$ values
are restricted. The fluid volume rate-of-expansion determines the type of
anisotropic solution. There are two possibilities: collapse with shrinking
volume and expansion with increasing volume.

The paper is organized as follows: section II describes the metric and metric
functions. Section III covers energy-momentum components and treats the
sectional curvature mass. Generated solutions are given in section IV. Section
V has matching conditions which relate the interior and exterior. The work is
summarized in section VI. Appendix A contains a general metric, an associated
energy-momentum tensor with heat flow, and trapped surface equations.

\section{METRIC AND COMPONENTS}

The collapse interior has an exterior which is vacuum Schwarzschild. Expanding
interiors have cosmological exteriors. Both collapse and expanding interiors
are covered by the spherically symmetric metric%
\begin{equation}
g_{\mu\nu}^{\text{aniso}}dx^{\mu}dx^{\nu}=A^{2}dt^{2}-B^{2}dr^{2}-R^{2}%
d\Omega^{2} \label{aniso-met}%
\end{equation}
where $A=A(t,r),B=B(t,r),R=R(t,r),$ and where $d\Omega^{2}$ is the metric of
the unit sphere.

The velocity is comoving with $\hat{u}^{\mu}\partial_{\mu}=A^{-1}\partial_{t}%
$. The interior metric is spanned by the tetrad
\begin{align*}
\hat{u}_{\mu}dx^{\mu}  &  =Adt,\ \ \hat{r}_{\mu}dx^{\mu}=Bdr,\\
\hat{\vartheta}_{\mu}dx^{\mu}  &  =Rd\vartheta,\ \ \hat{\varphi}_{\mu}dx^{\mu
}=R\text{sin}\vartheta d\varphi
\end{align*}
such that%
\begin{equation}
g_{\mu\nu}^{\text{aniso}}=\hat{u}_{\mu}\hat{u}_{\nu}-\hat{r}_{\mu}\hat{r}%
_{\nu}-\hat{\vartheta}_{\mu}\hat{\vartheta}_{\nu}-\hat{\varphi}_{\mu}%
\hat{\varphi}_{\nu}.
\end{equation}

\subsection*{Metric components}

Metric components are chosen so that the heat flow scalar $Q=0$. The equation
for $Q$ is written from Eq.(\ref{Q}), where primes denote $\partial/\partial
r$, and overdots denote $\partial/\partial t$, as%
\begin{equation}
\frac{A^{\prime}}{A}\frac{\dot{R}}{R}+\frac{\dot{B}}{B}\frac{R^{\prime}}%
{R}-\frac{\dot{R}^{\prime}}{R}=0. \label{zero-Q}%
\end{equation}
We find that,
\begin{align}
A  &  =h(t)\frac{\dot{R}}{R^{\alpha}},\text{ \ }h(t)\text{ arbitrary}%
\label{new-A}\\
B  &  =f(r)R^{\alpha},\text{ \ }f(r)\text{ arbitrary} \label{new-B}%
\end{align}
satisfies Eq.(\ref{zero-Q}) identically for arbitrary $R(t,r)$. As a metric
component $A=$ $h(t)\dot{R}/R^{\alpha}$ allows $h(t)$ to be absorbed by
redefining $dt$, similarly for $f(r)$ and $dr$. Therefore we set $h(t)=f(r)=1
$. The interior metric is now
\begin{equation}
g_{\mu\nu}^{\text{aniso}}dx^{\mu}dx^{\nu}=(\dot{R}/R^{\alpha})^{2}%
dt^{2}-R^{2\alpha}dr^{2}-R^{2}d\Omega^{2}. \label{aniso-met2}%
\end{equation}

Using values from Eqs.(\ref{new-A}) and (\ref{new-B}) for $A=\dot{R}%
/R^{\alpha}$ and $B=R^{\alpha}$, the fluid volume rate-of-expansion and
rate-of-shear scalar, given in Eq.(\ref{exp}) and Eq.(\ref{shear})
respectively, are%
\begin{align}
\Theta &  =(\alpha+2)R^{\alpha-1}\label{exp-value}\\
\sigma &  =(1-\alpha)R^{\alpha-1} \label{shear-val}%
\end{align}
thus restricting $\alpha:$ \ $\alpha\neq-2,1$. The value of $\Theta$ provides
two different interiors: collapse with $\alpha<-2$ and expansion with
$\alpha>1$.

We define a dimensionless measure of anisotropy as%
\begin{equation}
\bigtriangleup a:=\frac{p_{r}-p_{\perp}}{p_{r}}. \label{delta-a}%
\end{equation}

\section{ENERGY-MOMENTUM AND MASS}

The energy-momentum given in Appendix A is written here with zero heat flow
$q^{\mu}$, since the heat scalar is chosen to vanish%
\begin{equation}
T^{\mu\nu}=w\hat{u}^{\mu}\hat{u}^{\nu}+p_{r}\hat{r}^{\mu}\hat{r}^{\nu
}+p_{\perp}(\hat{\vartheta}^{\mu}\hat{\vartheta}^{\nu}+\hat{\varphi}^{\mu}%
\hat{\varphi}^{\nu})
\end{equation}
with fluid velocity $\hat{u}^{\mu}\hat{u}_{\mu}=1$ such that $\hat{u}^{\mu
}\partial_{\mu}=A^{-1}\partial_{t}$. $w$ is the mass-energy density, $p_{r}$
is the radial pressure and $p_{\perp}$ is the tangential pressure.

Substituting in equations (\ref{w}), (\ref{p-r}), and (\ref{p-perp}) for
$A=\dot{R}/R^{\alpha}$ and $B=R^{\alpha}$, we have%
\begin{equation}
8\pi w=(1+2\alpha)R^{2\alpha-2}-\frac{1}{R^{2\alpha}}\left[  2\frac
{R^{\prime\prime}}{R}+(1-2\alpha)(\frac{R^{\prime}}{R})^{2}\right]  +\frac
{1}{R^{2}} \label{w-eqn}%
\end{equation}%
\begin{equation}
8\pi p_{r}=\frac{1}{R^{2\alpha}}\left[  (2\alpha-1)(\frac{R^{\prime}}{R}%
)^{2}-2\frac{\dot{R}^{\prime}}{\dot{R}}(\frac{R^{\prime}}{R})\right]
+(1+2\alpha)R^{2\alpha-2}+\frac{1}{R^{2}} \label{pr-eqn}%
\end{equation}%
\begin{align}
8\pi p_{\perp}  &  =-\alpha(1+2\alpha)R^{2\alpha-2}\label{pperp-eqn}\\
&  +\frac{1}{R^{2\alpha}}\left[  \frac{\dot{R}^{\prime\prime}}{\dot{R}%
}+(1-3\alpha)\frac{\dot{R}^{\prime}}{\dot{R}}(\frac{R^{\prime}}{R}%
)+(1-\alpha)\frac{R^{\prime\prime}}{R}+\alpha(2\alpha-1)(\frac{R^{\prime}}%
{R})^{2}\right] \nonumber
\end{align}
Specifying $R(t,r)$ and $\alpha$ distinguishes an anisotropic configuration.

\subsection*{Sectional Curvature Mass}

Equation (\ref{sect-curv-mass}) provides the sectional curvature mass with
$A=\dot{R}/R^{\alpha}$ and $B=R^{\alpha}$%
\begin{equation}
\frac{2m}{R}=1+R^{2\alpha}-(R^{\prime})^{2}R^{-2\alpha} \label{trap-mass}%
\end{equation}
When $R^{\prime}=R^{2\alpha}$ there is a possible trapped surface at $R=2m$.
From equations (\ref{s1}) and (\ref{s2}) the trapping scalars are%
\begin{equation}
\kappa_{1}=\frac{R^{\alpha}}{R}+\frac{R^{\prime}}{R^{\alpha+1}},\text{
\ }\kappa_{2}=\frac{R^{\alpha}}{R}-\frac{R^{\prime}}{R^{\alpha+1}}%
\end{equation}
When $\kappa_{1}$ and $\kappa_{2}$ have the same sign, a trapped surface will
exist. For $R^{\prime}=R^{2\alpha}$ the trapping scalars are
\begin{equation}
\kappa_{1}=2R^{\alpha-1},\text{ \ }\kappa_{2}=0
\end{equation}
They are both non-negative, therefore during collapse a trapped surface
develops at $R=2m$. The trapped surface condition $R^{\prime}=R^{2\alpha}$ has
the integral
\begin{equation}
R_{\text{trap}}^{1-2\alpha}=(1-2\alpha)r+H(t),\ H\text{ arbitrary}%
\end{equation}

When the trapped surface condition $R^{\prime}=R^{2\alpha}$ is substituted
into equations (\ref{w-eqn},\ref{pr-eqn},\ref{pperp-eqn}) then
\begin{equation}
8\pi w=R_{\text{trap}}^{-2},\text{ \ }8\pi p_{r}=R_{\text{trap}}^{-2},\text{
\ }p_{\perp}\equiv0. \label{trapped-values}%
\end{equation}

\section{Generated Solutions}

\subsection*{(A) $\alpha=-5/2$}

The rate-of-expansion must be negative for collapse, therefore $\alpha$ must
be less than $-2$. Here we assume $\alpha=-5/2$. A trapped surface occurs when
$R^{\prime}=R^{2\alpha}$. Thus $R^{\prime}=R^{-5}$ or $(R^{6})^{\prime}%
=6$\ with solution%
\begin{equation}
R_{\text{trap}}=[6r+h_{1}(t)]^{1/6},\text{ \ }h_{1}(t)\text{ arbitrary}
\label{R-trap-soln}%
\end{equation}
The density and pressure equations for $\alpha=-5/2$ become%
\begin{align}
8\pi w  &  =-4R^{-7}-2R^{5}\left[  \frac{R^{\prime\prime}}{R}+3(\frac
{R^{\prime}}{R})^{2}\right]  +R^{-2}\\
8\pi p_{r}  &  =-2R^{5}\left[  3(\frac{R^{\prime}}{R})^{2}+(\frac{R^{\prime}%
}{R})\frac{\dot{R}^{\prime}}{\dot{R}}\right]  -4R^{-7}+R^{-2}\\
8\pi p_{\perp}  &  =-10R^{-7}+R^{5}\left[  \frac{\dot{R}^{\prime\prime}}%
{\dot{R}}+\frac{17}{2}\frac{\dot{R}^{\prime}}{\dot{R}}(\frac{R^{\prime}}%
{R})+\frac{7}{2}\frac{R^{\prime\prime}}{R}+15(\frac{R^{\prime}}{R}%
)^{2}\right]
\end{align}
We assume $R=k(6r+h_{1})^{1/6}.$ The density and pressures in this case are%
\begin{align}
8\pi w  &  =4k^{5}(1-k^{-12})(6r+h_{1})^{-7/6}+k^{-2}(6r+h_{1})^{-1/3}%
\label{w-1}\\
8\pi p_{r}  &  =4k^{5}(1-k^{-12})(6r+h_{1})^{-7/6}+k^{-2}(6r+h_{1}%
)^{-1/3}\label{pr-1}\\
8\pi p_{\perp}  &  =10(k^{4}-1)(6r+h_{1})^{-7/6}. \label{pperp-1}%
\end{align}
The mass-energy density $w(r)$ is graphed below in Figure 1.%

%TCIMACRO{\FRAME{fhFU}{4.0311in}{3.0366in}{0pt}{\Qcb{ $k=2,$ $h_{1}(t)=1,$ w vs
%r \ (unscaled)}}{}{w-vs-r-1.eps}{\special{ language "Scientific Word";
%type "GRAPHIC";  maintain-aspect-ratio TRUE;  display "USEDEF";
%valid_file "F";  width 4.0311in;  height 3.0366in;  depth 0pt;
%original-width 4.1556in;  original-height 3.1224in;  cropleft "0";
%filename 'w-vs-r-1.eps';file-properties "XNPEU";}}}%
%BeginExpansion
\begin{figure}[h]%
\centering
\includegraphics[
height=3.0366in,
width=4.0311in
]%
{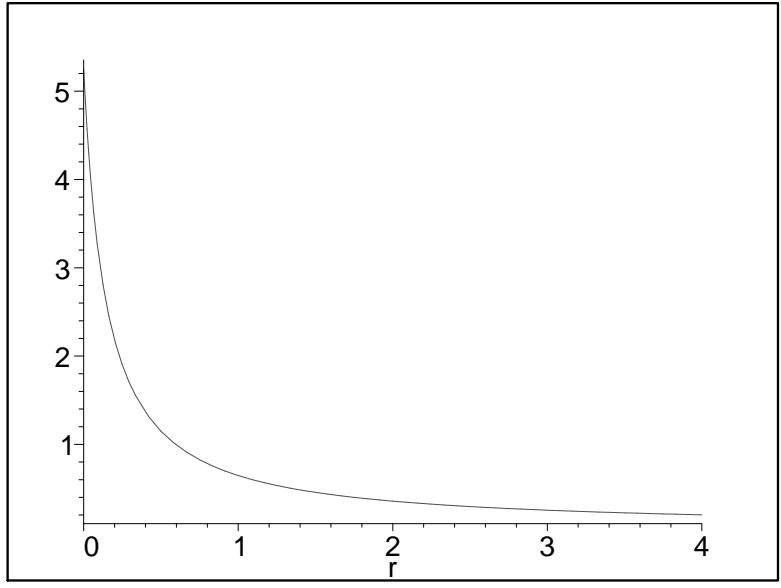}%
\caption{ $k=2,$ $h_{1}(t)=1,$ w vs r \ (unscaled)}%
\end{figure}
%EndExpansion
When $k=1$ the density and pressures revert to the trapped values given in
Eq.(\ref{trapped-values}). The sectional curvature mass given in
Eq.(\ref{trap-mass}) is%
\begin{equation}
2m_{1}(t,r)=R[1+(1-k^{2})R^{-5}] \label{mass1}%
\end{equation}
The constant $k$ and function $h_{1}(t)$ remain free choices.

The dimensionless measure of anisotropy, defined in Eq.(\ref{delta-a}), is
given by%
\begin{equation}
\bigtriangleup a=1+\frac{10k^{2}(1-k^{4})}{4k^{7}(1-k^{-12})+(6r+h_{1})^{5/6}}%
\end{equation}
For $h_{1}(t)$ constant and fixed $k$, the anisotropy is a maximum at the
center $r=0$. As $r$ increases $\bigtriangleup a$ falls to $1+\epsilon$ ($r$
has an upper bound - see the section on Matching). At the trapped surface,
where $k=1$, there is a minimum with $\bigtriangleup a=1$.

\subsection*{(B) $\alpha=3/2$}

We choose $\alpha=3/2$ with $\Theta=(5/2)R^{1/2}$ for expansion. We also
choose%
\begin{equation}
R=(r^{2}+r_{0}^{2})^{-1}+h_{2}(t),\text{ \ }h_{2}(t)\text{ arbitrary}%
\end{equation}
For $\alpha=3/2$
\begin{align}
8\pi w  &  =4R-2R^{-3}[\frac{R^{\prime\prime}}{R}-(\frac{R^{\prime}}{R}%
)^{2}]+R^{-2}\\
8\pi p_{r}  &  =2R^{-3}[(\frac{R^{\prime}}{R})^{2}-(\frac{R^{\prime}}{R}%
)\frac{\dot{R}^{\prime}}{\dot{R}}]+4R+R^{-2}\\
8\pi p_{\perp}  &  =-R[2\frac{\ddot{R}}{\dot{R}}(\frac{R}{\dot{R}}%
)+3]+R^{-3}[\frac{\dot{R}^{\prime\prime}}{\dot{R}}-\frac{3}{2}(\frac{\dot
{R}^{\prime}}{\dot{R}})(\frac{R^{\prime}}{R})-\frac{1}{2}\frac{R^{\prime
\prime}}{R}+\frac{3}{4}(\frac{R^{\prime}}{R})^{2}]
\end{align}
With $F(t,r):=1+h_{2}(t)(r^{2}+r_{0}^{2})$ and $R=F/(r^{2}+r_{0}^{2})$, the
density and pressures in this case are%
\begin{align}
8\pi w  &  =\frac{4F}{r^{2}+r_{0}^{2}}+\frac{(r^{2}+r_{0}^{2})^{2}}{F^{2}%
}+\frac{4(r^{2}+r_{0}^{2})(r_{0}^{2}-3r^{2})}{F^{4}}+\frac{8r^{2}(r^{2}%
+r_{0}^{2})}{F^{5}}\\
8\pi p_{r}  &  =\frac{4F}{r^{2}+r_{0}^{2}}+\frac{(r^{2}+r_{0}^{2})^{2}}{F^{2}%
}+\frac{8r^{2}(r^{2}+r_{0}^{2})}{F^{5}}\\
8\pi p_{\perp}  &  =-\frac{F}{r^{2}+r_{0}^{2}}[2F\frac{\ddot{F}}{(\dot{F}%
)^{2}}+3]+\frac{(r^{2}+r_{0}^{2})^{2}(r_{0}^{2}-3r^{2})}{F^{4}}-\frac{3r^{2}%
}{F^{2}(r^{2}+r_{0}^{2})^{2}}%
\end{align}
%

%TCIMACRO{\FRAME{dhFU}{4.0311in}{3.0366in}{0pt}{\Qcb{Fig.2 $\ h_{2}(t)=1$,
%$r_{0}=1$, \ w vs r \ (unscaled)}}{}{w-vs-r-2.eps}%
%width 4.0311in;  height 3.0366in;  depth 0pt;  original-width 4.1556in;
%original-height 3.1224in;  cropleft "0";  croptop "1";  cropright "1";
%cropbottom "0";  filename 'w-vs-r-2.eps';file-properties "XNPEU";}}}%
%BeginExpansion
\begin{center}
\includegraphics[
height=3.0366in,
width=4.0311in
]%
{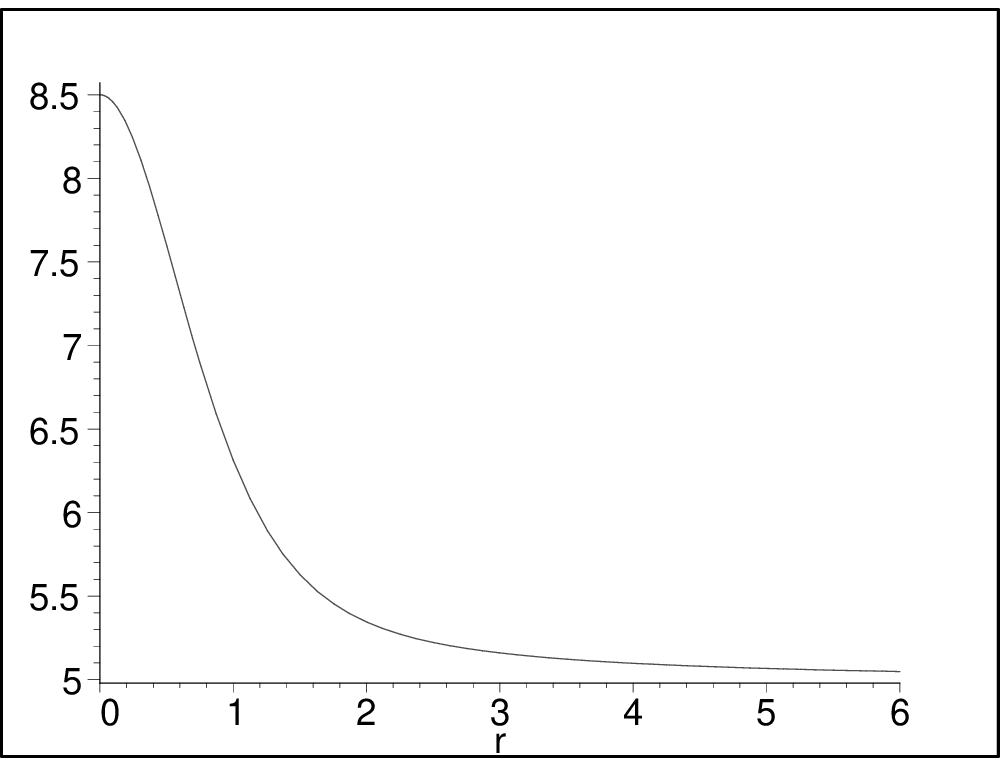}%
\\
Fig.2 $\ h_{2}(t)=1$, $r_{0}=1$, \ w vs r \ (unscaled)
\end{center}
%EndExpansion
The sectional curvature mass is%
\begin{equation}
2m_{2}(t,r)=\frac{F}{r^{2}+r_{0}^{2}}\left[  1+\frac{F^{3}}{(r^{2}+r_{0}%
^{2})^{3}}-\frac{4r^{2}}{(r^{2}+r_{0}^{2})F^{3}}\right]  .
\end{equation}
When $h_{2}(t)=1$ and $r_{0}=1$, as in Figure 2, then as $r\rightarrow\infty$
\ $2m_{2}\rightarrow$ $2$.

\section{Matching}

Interior and exterior regions match across a separating boundary $\Sigma$ if
the first and second fundamental forms are continuous across $\Sigma$
\cite{MS93}. The collapsing interiors match to a Schwarzschild exterior
\begin{equation}
ds_{(+)}^{2}=(1-2m_{0}/\tilde{r})d\tilde{t}^{2}-(1-2m_{0}/\tilde{r}%
)^{-1}d\tilde{r}^{2}-\tilde{r}^{2}d\Omega^{2}. \label{Schw-met}%
\end{equation}
The interior metric is
\begin{equation}
ds_{(-)}^{2}=A^{2}dt^{2}-B^{2}dr^{2}-R^{2}d\Omega^{2}. \label{int-met}%
\end{equation}
(Interior objects will be indicated by a minus and exterior objects by a
plus.) The metric match requires%
\begin{align}
A  &  =(1-2m_{0}/r_{\Sigma})^{1/2}\\
B  &  =(1-2m_{0}/r_{\Sigma})^{-1/2}\\
R  &  =r_{\Sigma}%
\end{align}
At the boundary, the sectional curvature mass must equal the exterior
Schwarzschild mass parameter%
\begin{equation}
m(r_{\Sigma})=m_{0} \label{match2}%
\end{equation}
The unit normal to the boundary is%
\begin{equation}
N_{\mu}dx^{\mu}=-(1-2m_{0}/r_{\Sigma})^{1/2}d\tilde{t}+\sqrt{2}(1-2m_{0}%
/r_{\Sigma})^{-1/2}d\tilde{r}%
\end{equation}
such that $g_{(+)}^{\mu\nu}N_{\mu}N_{\nu}=-1$

Junction conditions equivalent to continuity of the second fundamental form
are
\begin{equation}
(G_{\mu\nu}N^{\mu}N^{\nu})_{^{\Sigma}}^{+}=(G_{\mu\nu}N^{\mu}N^{\nu
})_{^{\Sigma}}^{-}, \label{G-N-N}%
\end{equation}%
\begin{equation}
(G_{\mu\nu}N^{\mu}e_{^{j}}^{\nu})_{^{\Sigma}}^{+}=(G_{\mu\nu}N^{\mu}e_{^{j}%
}^{\nu})_{^{\Sigma}}^{-}, \label{G-N-tet}%
\end{equation}
where three linearly independent $e_{^{j}}^{\nu}$ span the boundary surface.
The Schwarzschild exterior is vacuum, so that%
\begin{align*}
(G_{\mu\nu}N^{\mu}N^{\nu})_{^{\Sigma}}^{-}  &  =0,\\
(G_{\mu\nu}N^{\mu}e_{^{j}}^{\nu})_{^{\Sigma}}^{-}  &  =0.
\end{align*}
This implies that at the boundary
\begin{equation}
(w)_{\Sigma}=(p_{r})_{\Sigma}=(p_{\perp})_{\Sigma}=0
\end{equation}
i.e. the density and pressures must vanish. Figure 1 show the density falling
off to zero. The pressures fall off similarly.

Expanding solutions have cosmological exteriors. Collins \cite{Col77}
discussed the global properties of Kantowski-Sachs \cite{KS66} cosmologies. He
dropped their dust condition and showed in detail how the expanding interiors
relate to their cosmological exteriors.

\section{SUMMARY}

Analytic anisotropic collapse and expansion solutions have been generated. The
interior collapse metric has been matched to an exterior Schwarzschild vacuum.
The collapse solution has a trapped surface where the dimensionless anisotropy
measure is minimized. To paraphrase Herrera and Santos \cite{HS97}: "For dense
systems, phase transitions may occur during gravitational collapse,
particularly transitions to a pion condensed state. A softened equation of
state can provide a large energy release which is important in the evolution
of collapsing configurations." Additionally, redshift data from the Coma
cluster of galaxies \cite{GDK99} shows a distinct infall region.

Expanding solutions have been discussed by Collins \cite{Col77}. These
solutions allow inhomogeneous cosmologies to be modelled. For instance, Barrow
and Maartens \cite{BM98} have investigated the effect of anisotropic stresses
on the late-time evolution of inhomogeneous universes. They found decay of
shear anisotropy to be slowed by the presence of anisotropic stresses. They
also found attractor solutions relating distortion in the expansion anisotropy
to the fractional density in anisotropic stress.

This work differs from previous studies such as
\cite{HOP08,BL74,CHE+81,DG02,MH03} which are static, and studies \cite{HPO01}
in which the spherical area radius is simplified to $r$, rather than $R(t,r)$
used here. We find this is the only work which uses $R(t,r)$ to generate
solutions. In summary, the solution generating function is $R(t,r)$ with
parameter "$\alpha$". $R(t,r)$ was constructed by requiring zero heat flow
scalar. For each valid choice of $\alpha$ and $R(t,r)$, one obtains an
anisotropic configuration which collapses or expands, and each generating
function contains an arbitrary function of time which can be chosen to fit
various astrophysical time profiles.

\begin{center}
\textbf{ACKNOWLEDGMENT}

We thank Professor Jean Krisch for constructive comments.
\end{center}

\appendix

\section{Energy-Momentum and physical components}

Metric
\[
g_{\mu\nu}^{\text{aniso}}dx^{\mu}dx^{\nu}=A^{2}dt^{2}-B^{2}dr^{2}-R^{2}%
d\Omega^{2}%
\]
is Petrov type \textbf{D. }The two principal null vectors, normal to
($\vartheta,\varphi$) two-surfaces, are%
\begin{align}
l^{\mu}\partial_{\mu}  &  =A^{-1}\partial_{t}+B^{-1}\partial_{r}%
\label{l-vec}\\
n^{\mu}\partial_{\mu}  &  =A^{-1}\partial_{t}-B^{-1}\partial_{r}.
\label{n-vec}%
\end{align}
The energy-momentum tensor is given by ($G=c=1$)
\begin{equation}
T^{\mu\nu}=w\hat{u}^{\mu}\hat{u}^{\nu}+p_{r}\hat{r}^{\mu}\hat{r}^{\nu
}+p_{\perp}(\hat{\vartheta}^{\mu}\hat{\vartheta}^{\nu}+\hat{\varphi}^{\mu}%
\hat{\varphi}^{\nu})+q^{\mu}\hat{u}^{\nu}+\hat{u}^{\mu}q^{\nu} \label{en-mom}%
\end{equation}
where $p_{r}$ is the radial pressure, $p_{\perp}$ is the tangential pressure,
$w$ is the mass-energy density, and $q^{\mu}$ is the radial heat flow vector
orthogonal to $\hat{u}^{\mu}$. We use notation of Taub \cite{Tau69} for $w$.
Taub's purpose was to distinguish mass-energy density from proper mass-density
$\rho$, with $w=\rho(1+\epsilon)$. This allows the first law of thermodynamics
to be written in its usual form $Tds=d\epsilon+pd(1/\rho)$ with specific
entropy $s$ and specific internal energy $\epsilon$. The kinematics of the
fluid are described by
\begin{subequations}
\label{three}%
\begin{align}
\hat{u}_{\mu;\nu}  &  =a_{\mu}\hat{u}_{\nu}+\sigma_{\mu\nu}-(\Theta/3)(\hat
{r}_{\mu}\hat{r}_{\nu}+\hat{\vartheta}_{\mu}\hat{\vartheta}_{\nu}+\hat
{\varphi}_{\mu}\hat{\varphi}_{\nu}),\\
a_{\mu}dx^{\mu}  &  =-(A^{\prime}/A)dr,\\
\Theta &  =A^{-1}(\dot{B}/B+2\dot{R}/R),\label{exp}\\
\sigma_{\mu\nu}  &  =\sigma\lbrack-2\hat{r}_{\mu}\hat{r}_{\nu}+\hat{\vartheta
}_{\mu}\hat{\vartheta}_{\nu}+\hat{\varphi}_{\mu}\hat{\varphi}_{\nu
}],\ \ \sigma=A^{-1}(\dot{R}/R-\dot{B}/B) \label{shear}%
\end{align}
where primes denote $\partial/\partial r$, and overdots denote $\partial
/\partial t$. The rate-of shear $\sigma_{\mu\nu}$ is trace-free, and
$\sigma_{\mu\nu}\sigma^{\mu\nu}=6\sigma^{2}$. The heat flow vector ($q^{\mu
}\hat{u}_{\mu}=0$) is given by
\end{subequations}
\begin{align}
4\pi q^{\mu}\partial_{\mu}  &  =Q\partial_{r}\label{q-vec}\\
Q  &  =(AB)^{-1}\left[  \frac{A^{\prime}}{A}\frac{\dot{R}}{R}+\frac{\dot{B}%
}{B}\frac{R^{\prime}}{R}-\frac{\dot{R}^{\prime}}{R}\right]  \label{Q}%
\end{align}
The sectional curvature mass is
\begin{equation}
2m=R^{3}R_{\alpha\beta\mu\nu}\hat{\vartheta}^{\alpha}\hat{\varphi}^{\beta}%
\hat{\vartheta}^{\mu}\hat{\varphi}^{\nu}=R[1+\dot{R}^{2}/A^{2}-(R^{\prime
})^{2}/B^{2}]. \label{sect-curv-mass}%
\end{equation}
The mass-energy density and pressures are given, respectively, by%
\begin{equation}
8\pi w=\frac{1}{A^{2}}\left[  (\frac{\dot{R}}{R})^{2}+2\frac{\dot{R}}{R}%
\frac{\dot{B}}{B}\right]  -\frac{1}{B^{2}}\left[  2\frac{R^{\prime\prime}}%
{R}+(\frac{R^{\prime}}{R})^{2}-2\frac{B^{\prime}}{B}\frac{R^{\prime}}%
{R}\right]  +\frac{1}{R^{2}} \label{w}%
\end{equation}%
\begin{equation}
8\pi p_{r}=\frac{1}{B^{2}}\left[  (\frac{R^{\prime}}{R})^{2}+2\frac{R^{\prime
}}{R}\frac{A^{\prime}}{A}\right]  -\frac{1}{A^{2}}\left[  2\frac{\ddot{R}}%
{R}+(\frac{\dot{R}}{R})^{2}-2\frac{\dot{R}}{R}\frac{\dot{A}}{A}\right]
-\frac{1}{R^{2}} \label{p-r}%
\end{equation}%
\begin{equation}
8\pi p_{\perp}=-\frac{1}{A^{2}}[\frac{\ddot{R}}{R}+\frac{\ddot{B}}{B}%
+\frac{\dot{R}}{R}(\frac{\dot{B}}{B}-\frac{\dot{A}}{A})+\frac{\dot{A}}{A}%
\frac{\dot{B}}{B}]+\frac{1}{B^{2}}[\frac{R^{\prime\prime}}{R}+\frac
{A^{\prime\prime}}{A}+\frac{R^{\prime}}{R}(\frac{A^{\prime}}{A}-\frac
{B^{\prime}}{B})-\frac{A^{\prime}}{A}\frac{B^{\prime}}{B}] \label{p-perp}%
\end{equation}

\subsection*{Trapped Surfaces}

The topological two-spheres ($\vartheta,\varphi$) nested in an $R=const$
surface at time $t$ have outgoing null geodesic normal $l^{\mu}$ and incoming
null geodesic normal $n^{\mu}$. The two principal null vectors (\ref{l-vec})
and (\ref{n-vec}) provide trapping scalars%
\[
\kappa_{1}=l^{\mu}\partial_{\mu}(\ln R),\text{ \ }\kappa_{2}=n^{\mu}%
\partial_{\mu}(\ln R).
\]
When scalars \cite{Sen02} $\kappa_{1}$ and $\kappa_{2}$ have the same sign, a
trapped surface \cite{Pen65} will exist:
\begin{align}
\kappa_{1}  &  =A^{-1}(\dot{R}/R)+B^{-1}(R^{\prime}/R)\label{s1}\\
\kappa_{2}  &  =A^{-1}(\dot{R}/R)-B^{-1}(R^{\prime}/R). \label{s2}%
\end{align}

\end{document}